# Magnetoplasmonics in nanocavities: Dark plasmons enhance magneto-optics beyond the intrinsic limit of magnetoplasmonic nanoantennas

*Alberto López-Ortega*[1,*], *Mario Zapata Herrera*[1], *Nicolò Maccaferri*[2], *Matteo Pancaldi*[3], *Mikel Garcia*[1], *Andrey Chuvilin*[1,4], *and Paolo Vavassori*[1,4,*]

[1] CIC nanoGUNE, Donostia–San Sebastian 20018, Spain

[2] Physics and Materials Science Research Unit, Université du Luxembourg, L-1511 Luxembourg, Luxembourg

[3] Department of Physics, Stockholm University, 106 91 Stockholm, Sweden

[4] IKERBASQUE, Basque Foundation for Science, Bilbao 48013, Spain

*Correspondence to: p.vavassori@nanogune.eu; lopezortega.alberto@gmail.com;

**Enhancing magneto-optical effects is crucial for size reduction of key photonic devices based on non-reciprocal propagation of light and to enable active nanophotonics. We disclose a so far unexplored approach that exploits dark plasmons to produce an unprecedented amplification of magneto-optical activity. We designed and fabricated non-concentric magnetoplasmonic-disk/plasmonic-ring-resonator nanocavities supporting multipolar dark modes. The broken geometrical symmetry of the design enables coupling with free-space light and hybridization of dark modes of the ring nanoresonator with the dipolar localized plasmon resonance of the magnetoplasmonic disk. Such hybridization generates a multipolar resonance that amplifies the magneto-optical response of the nanocavity by ~1-order of magnitude with respect to the maximum enhancement achievable by localized plasmons in bare magnetoplasmonic nanoantennas. This large amplification results from the peculiar and enhanced electrodynamic response of the nanocavity, yielding an intense magnetically-activated *radiant magneto-optical dipole* driven by the *low-radiant* multipolar resonance. The concept proposed is general and, therefore, our results open a new path that can revitalize research and applications of magnetoplasmonics to active nanophotonics and flat optics.**

Nanophotonics uses light polarization as an information carrier in optical communications, sensing, and imaging.[1] Likewise, the state of polarization plays a key role in the photonic transfer of quantum information.[2] In this framework, optical nanocomponents enabling dynamic manipulation of light polarization at the nanoscale are key features for future nanophotonic applications. Electromagnetic interaction with plasmonic meta-surfaces and crystals has emerged as a prominent route to develop more efficient devices for the active control of light at small scales.[3–6] The development of these artificial materials, dubbed metamaterials, has been tightly linked with the current advances in the nanofabrication allowing the preparation of extremely sub-wavelength nanostructures, which combine diverse optical properties.[5] One relevant example is the class of magnetoplasmonic surfaces and crystals, composed by arrangements of nanoantennas either entirely[7–13] or partially[14–20] made of magnetic materials. Magnetic materials exhibit the so-called magneto-optical (MO) activity, arising from spin-orbit coupling of electrons, which results in a weak magnetic-field-induced intensity and polarization modulation of reflected and transmitted light. The unique optical properties of magnetoplasmonic nanoantennas arise from combining strong local enhancements of electromagnetic fields via surface plasmon excitations with their inherent MO activity. In the past decade, materials based on magnetoplasmonic nanoantennas were intensively investigated for their non-reciprocal light propagation properties, aiming for 2D flat-optics nanodevices, such as rotators, modulators, and isolators.[9,11,14–24] Their ability to enhance optical and MO responses resulted in an ultimate accuracy in the measurement of distances at the nanoscale[25] as well as very small refractive index changes in label-free biosensing applications.[8,12,26]

Up to now, most studies of magnetoplasmonic nanostructures utilized bright (radiative) plasmon modes, such as localized dipolar plasmonic resonances (LPRs). Indeed, multilayered hybrid noble/ferromagnetic metal structures and purely ferromagnetic nanoantennas have demonstrated the possibility to control and amplify the MO response via bright plasmons excitation.[7–20] For a circular disk-like magnetoplasmonic nanoantenna, incident radiation of proper wavelength excites a LPR. Due to the activation by external magnetic field $\boldsymbol{H}$ of a net magnetization $\boldsymbol{M}$ in a ferromagnetic nanoantenna, a second LPR is induced by the inherent MO activity.[27] The MO-induced LPR (MOLPR) is driven by the optical LPR in a direction orthogonal to both $\boldsymbol{M}$ and the LPR itself. The ratio of the amplitudes and the phase lag between these two orthogonal resonant bright electric dipoles determines the magnetic-field-induced polarization change of interacting light.[27] For typical metallic constituents (pure ferromagnets as well as hybrid noble-metals/ferromagnets), both LPR and MOLPR of magnetoplasmonic nanoantennas have a relatively low quality factor (Q-factor). A maximum Q-factor in the order of 10 is typical of Au and Ag nanoantennas in the visible-near infrared spectral range and this value can be considered the upper limit also for magnetoplasmonic nanoantennas. The amplitude of the electric dipole associated to the LPR is approximately Q times, i.e, at most *up to* ~1-order of magnitude, larger than that induced in the continuous film counterpart. In a disk-shapes nanoantenna, the amplitude of the corresponding MOLPR driven by the LPR would be approximately $Q^2$ times, i.e., *up to* ~2-orders of magnitude larger than that in the continuous film in the best-case scenario considered here. However, the MO activity is proportional to the ratio between MOLPR and LPR amplitudes.[27] Therefore the resonant excitation of the two concurrent bright electric dipoles associated to the MOLPR and LPR limits the maximum achievable enhancement of the MO activity to only up to ~1-order of magnitude as it was observed in previous experiments.[7–20] This *up to 1-order of magnitude* enhancement of the MO response represents therefore a sort of fundamental upper limit achievable with metallic magnetoplasmonic nanoantennas.

Recently the possibility to excite multipolar dark modes in complex plasmonic structures has emerged within the nanophotonics community as an important topic[28–40] due to their impact on nanoscaled lasing effects,[41] plasmon-driven strong-coupling dynamics,[42] hot-electron generation,[43] and ultrasensitive molecular detection.[30] Archetypical structures investigated are symmetric nanorings and concentric ring/disk nanocavities. For concentric ring/disk nanocavities, electromagnetic radiation at normal incidence can excite only dipolar modes in the ring (so-called bonding and antibonding modes)[34–37] that couple to the dipolar mode of the disk leading to subradiant and superradiant coupled modes. Subradiant modes have been extensively studied in the last years due to their richer optical behavior and its potential implication for refractive index sensing,[38] enhanced spectroscopy,[32,34] and nanolasers.[40] Alike bare ring nanoresonators, concentric ring/disk nanocavities can sustain multipolar dark modes, but their charge distribution forbids a direct coupling to free-space photons at normal incidence. However, excitation of dark modes is enabled when the rotational symmetry of the unit is broken, e.g., by displacing the disk position away from the ring center (non-concentric ring/disk nanocavities), leading to the appearance of new modes corresponding to Fano interferences generated by the hybridization of the dipolar mode of the disk and the multipolar dark modes of the ring.[32–34]

The combination of magnetoplasmonics and nanocavities is an unexplored terrain. Up to now, plasmonic nanocavities showing the excitation of multipolar resonances have been synthesized from non-magnetic plasmonic materials, typically gold.[29–34,37] In this letter we propose a design and demonstrate the fabrication of bi-component magnetoplasmonic nanocavities composed of a gold ring resonator and a ferromagnetic disk asymmetrically placed inside the ring. The magnetoplasmonic functionality of this structure stems from the excitation of a Fano resonance resulting from the hybridization of the LPR in the magnetic disk with a multipolar dark mode in the plasmonic ring. This hybridized mode features an intense bonding character but results to be *low-radiant* and thus does not significantly enhance the reemission of the light with primary polarization. In turn, when the magnetic nanodisk is magnetically activated under the application of a magnetic field $\boldsymbol{H}$ parallel to the wavevector of the incident electromagnetic wave, the hybrid *low-radiant* mode drives an intense and bright, i.e. *radiant*, MOLPR mode in the ferromagnetic nanodisk, which is not hybridized. The MOLPR results being much more intense (~1-order of magnitude) than the enhanced MOLPR achievable in bare magnetoplasmonic nanoantennas. We explore the unique potential of such construct for enhancing and controlling the polarization of re-emitted light via modulation of an external magnetic field.

It is worth noting that the nanofabrication of multi-component structures combining different materials in non-simply connected geometries, like non-concentric ring/disk nanocavities fabricated here, require a positioning control of individual parts of ~10 nm. This represents an extremely challenging requirement even for modern nanofabrication. Only recent studies reported on the fabrication of nanorings and split-ring resonators combining plasmonic and magnetic materials, but with the latter only as part of the ring, via colloidal lithography.[24,44]

## Results and discussion

A schematic of the NCRD nanocavity together with atomic force and scanning electron microcopy images of an array of such nanocavities are depicted in Figs. 1a-c. The Au nanocavity is characterized by inner and outer ring radii of $R_i$ = 130 nm and $R_o$ = 215 nm, respectively; the radius of the Py disk is $R_d$ = 50 nm, and the gap between the disk and the ring is $g \approx 10$ nm. For comparison, arrays of isolated Py disks (Py-DI) magnetoplasmonic nanoantennas and isolated Au rings (Au-RI) have also been fabricated for reference and are depicted in Figs. 1d-e. We also fabricated a control sample with an array of concentric ring/disk nanocavities using the same $R_i$, $R_o$, and $R_d$ (Fig. S6). All the structures are 40-nm-thick and are arranged in a square array of 780-nm-pitch. The synthesis has been carried out by electron-beam lithography, followed by electron-beam and thermal evaporation of the Py and Au materials, respectively, onto Pyrex substrates. Specifically, the synthesis of the hybrid Py/Au nanocavities has been achieved by finely controlled two-step electron-beam lithography process in order to grow sequentially the Au rings in the desired position around the preexisting Py disks. We stress here that former studies of non-concentric ring/dot nanocavities dealt with mono-component structures typically made of Au, due to the high demands that their synthesis poses to nanolithography.

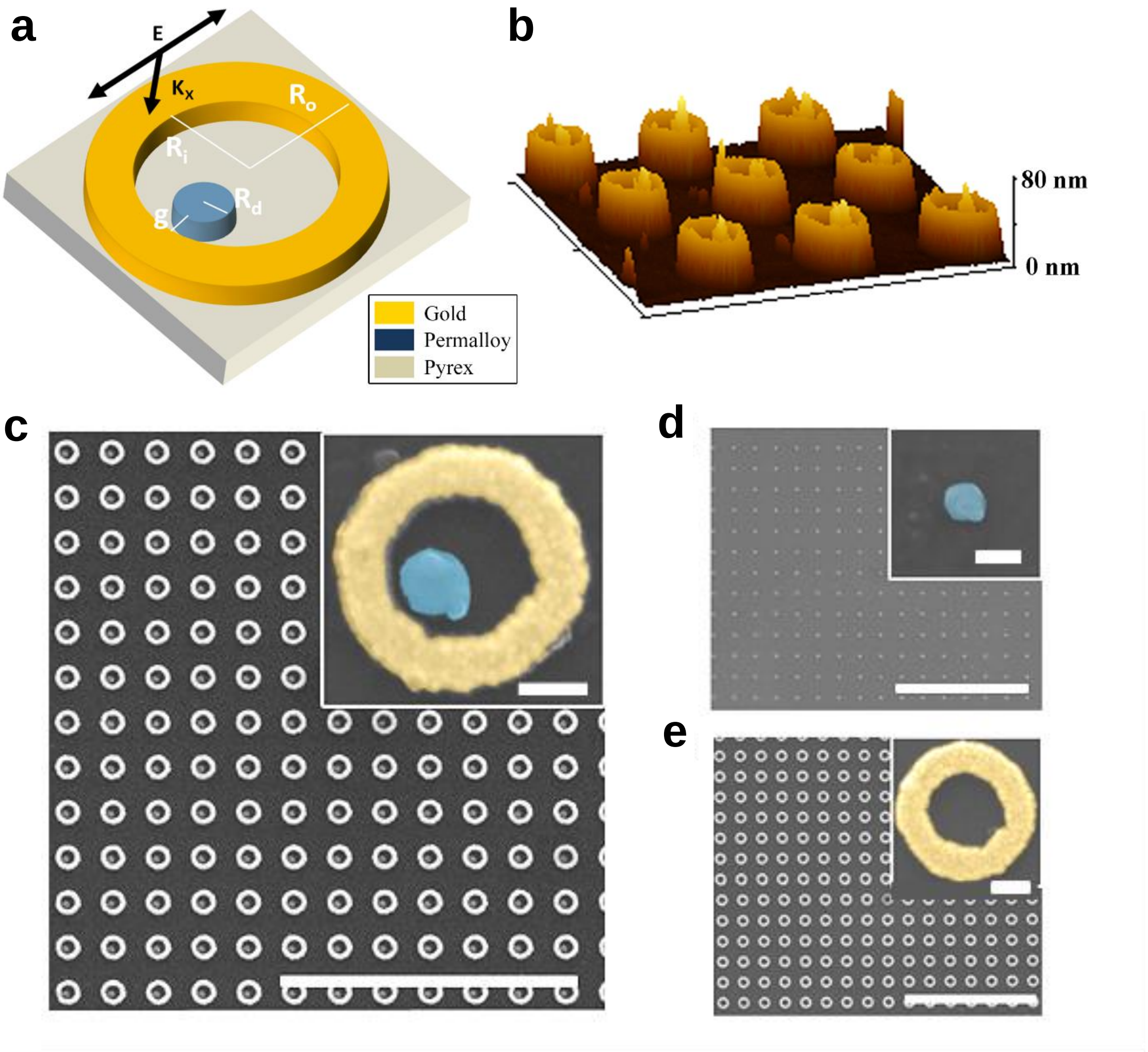

**Fig. 1 Magnetoplasmonic NCRD ferromagnetic-nanoantenna/gold-nanocavity and parent Py-DI and Au-RI nanostructures. a** Schematic of the NCRD hybrid structure with its four geometric characteristic parameters. **b** Atomic force and **c** SEM images of the individual NCRD nanocavity and array. SEM images of the parent single **d** Py-DI, **e** Au-RI constituents and arrays. The scale bars in panels **c-e** correspond to 5 µm and those in their insets to 100 nm.

Simulated and experimental transmittance spectra for the studied structures are depicted in Figs. 2a-b. Both experimental and simulated spectra of the NCRD array display two strongly marked dips located at 600 and 1650 nm and a weaker dip at 820 nm. A comparison with the spectra (simulated and experimental) of the array of bare Au-RI and a close inspection of the spectral dependence of calculated surface charge distribution maps in Fig. 2c, reveal that the two most prominent dips in the NCRD correspond to the excitation of the so-called antibonding and bonding plasmonic resonances in the Au ring portion of the nanocavity at 600 nm and 1650, respectively (see also Fig. S1).[36,37] As featured in Figs. 2c and S1, the lower wavelength antibonding resonance corresponds to a dipolar mode through the inner and the outer surface of the ring; the bonding mode at 1600 nm corresponds to a dipolar resonance involving the entire ring structure. Both resonances are bright modes for the Au-RI that can be excited by direct coupling with free-space light even in symmetric structures, and thus appear in both the NCRD and Au-RI spectra. The presence of the Py disk nanoantenna in the NCRD is only marginally perturbing these modes, which occur at roughly the same wavelength and with almost identical features in both the NCRD and Au-RI spectra. For the NCRD, these two resonances correspond to the superradiant (600 nm) and subradiant (1650 nm) modes reported in the literature for mono-component cavities.[32] It is worth noting that far-field diffractive coupling due to the periodic array design of the samples produces extremely weak features at 800 nm in the simulated spectrum of AuRI (coupling through air, small black arrow in Fig. 2a).[28] Such features are not observed in the experimental spectra and therefore far-field diffractive coupling can be regarded as negligible for our samples. The array with standalone Py disks produces a very broad plasmonic dipolar resonance peaked at ~550 nm (black solid lines in Figs. 2a-b).[7]

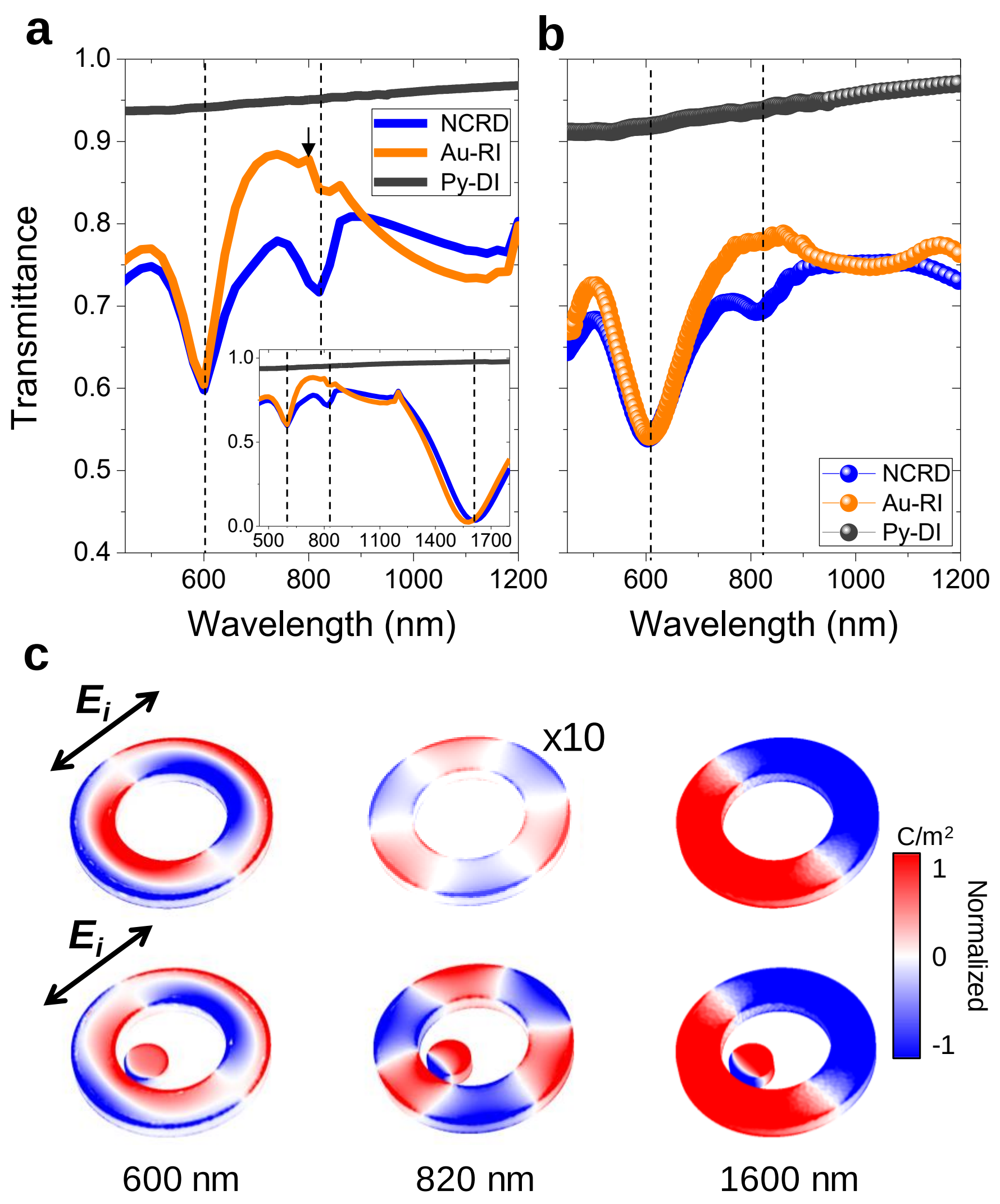

**Fig. 2 Au-RI and NCRD optical properties and electrodynamics. a** Simulated and **b** experimental transmittance spectra for the NCRD, Py-DI and Au-RI structures. Dashed lines mark the major features in the spectra at 600, 820 and 1600 nm. The small black arrow in panel **a** highlights a minor feature due to the weak far-field diffractive coupling in a simulated periodic array of defect-less structures. **c** Surface charge density maps (see Methods) for the Au-RI and NCRD structures at 600, 820 and 1600 nm, normalized to the map at 820 nm for the NCRD for direct comparison. Simulations in panel **c** are carried out using linearly polarized electromagnetic radiation as indicated by the black arrow ($E_i$ = 1V/m). The surface charge density for the Au-RI at 820 nm has been multiplied by a factor 10 for visualization purposes.

A comparison of surface charge distribution maps of the isolated Au-RI and the NCRD nanocavity at 820 nm (Fig. 2c) clearly reveals that the dip at this wavelength arises from the strong and localized near-field coupling between the broad dipolar resonance of the Py disk with a high-order multi-polar dark mode possessing a $S_6$ reflection-rotational (6-fold improper rotation) symmetry in the Au ring portion of the nanocavity. In literature, this mode is referred to as either the hexapolar or the octupolar mode of a ring.[32–34] Hereafter, we will label this mode as $\boldsymbol{S_6}$ in reference to the point-group notation. Inspection of Fig. 2c shows that the surface charge distribution of the Au-RI displays a very weak $\boldsymbol{S_6}$ at 820 nm (note that intensity of the surface charge density map had to be multiplied by a factor 10 to become visible in Fig. 2c), while for the NCRD the excitation of an intense $\boldsymbol{S_6}$ mode is clearly visible, although it is slightly distorted due to hybridization with the dipole in the Py disk.

To confirm experimentally a complete plasmonic spectrum of gold cavities we have performed an STEM-EELS study. The localized excitation realized in STEM-EELS,[45] can indeed efficiently excite all the eigenmodes supported by a plasmonic structure, including non-radiative modes, however, due to the rotational symmetry of the rings, a spatial distribution of the modes cannot be visualized. For this purpose, we have fabricated Au nanorings with the same dimensions as those of our NCRD nanocavities on a 20-nm-thick $SiN_x$ membrane. Complementary, we conducted a detailed analytical calculation of all possible plasmonic resonances that Au ring structure can support in the spectral range 500-2000 nm (0.6 - 2.5 eV) (a quasi-normal-mode expansion formalism[46] was used, see Methods for details). A comparison of the results is presented in Fig. S1. Additionally, there are two multipolar dark modes with energies between those of the antibonding and bonding modes. These modes are the $\boldsymbol{S_6}$ mode at around 775 nm (~1.6 eV) and the quadrupolar $\boldsymbol{S_4}$ mode at longer wavelength, slightly above 1000 nm (~1.2 eV). The EELS spectrum shown in Fig. S1b, which is an average of spectra taken on different Au-RIs to account for size distribution and nanofabrication defects, displays 4 clear peaks, one broad and centered at 2.1 eV (590 nm, antibonding mode) followed by narrower peaks centered at 1.55 eV (800 nm, $\boldsymbol{S_6}$ mode), 1.2 eV (1050 nm, $\boldsymbol{S_4}$ quadrupolar mode), and 0.72 eV (1720 nm, dipolar bonding mode) in excellent agreement with the predictions of our numerical simulations and analytical calculations.

In our NCRD nanocavity, a coupling of the dipolar plasmon resonance of the Py disk nanoantenna can only occur with the strong bright dipolar antibonding mode and with the $\boldsymbol{S_6}$ dark mode at around 600 nm and 800 nm of wavelength, respectively. We therefore expect to see only three features in the spectrum of NCRD structures, at around wavelengths of 600 nm, 800 nm, and above 1600 nm, which is exactly what we observe in the simulated and measured spectra (Figs. 2a and 2b).

Once we attained a clear overview of the main optical features, we then investigated the physics of the hybridization of bright and dark modes in the magnetoplasmonic nanocavity and its effects on the $\boldsymbol{H}$-induced light polarization modulation in the relevant spectral range, i.e., from 500 nm to 1200 nm where the hybridization should occur. The study was performed by measuring MO Kerr effect (MOKE) spectra, namely polarization rotation ($\theta_K$) and ellipticity ($\varepsilon_K$), of reflected light while changing magnetization of the Py disk with $\boldsymbol{H}$ applied perpendicular to the sample plane. Figure 3a shows a schematic of the MOKE configuration utilized in the experiment (polar MOKE configuration, see Methods for details). From the spectra of $\theta_K$ and $\varepsilon_K$, the spectral dependence of the $\boldsymbol{H}$-induced modulation of light polarization is numerically quantified by the modulus $|\Theta_K|$ of the complex Kerr angle, $\Theta_K = (\theta_K + i\ \varepsilon_K)$. This quantity is conventionally named as MO activity, hereafter referred to as MOA. The measured $\theta_K$, $\varepsilon_K$, and MOA spectra are reported in Figs. 3b and 3c for both the Py-DI and the NCRD structures.

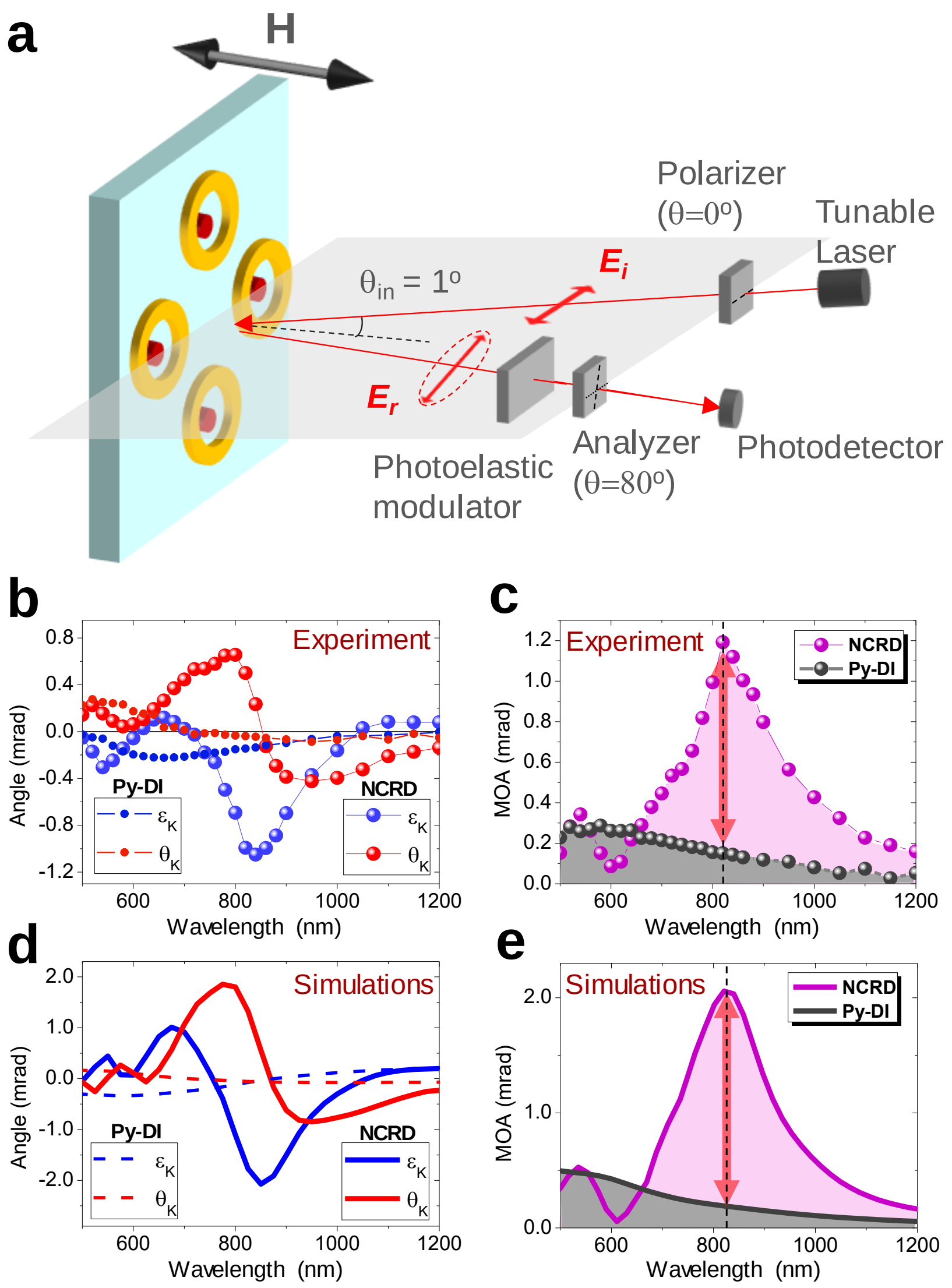

**Fig. 3 Magneto-optical response of the NCRD hybrid structure and the bare Py nanoantenna. a** Schematic of the magneto-optical Kerr effect spectrometer. The setup consists of a broadband supercontinuum tunable laser, polarizing and focusing optics, a photoelastic modulator, and a photodetector. The instrument is operated under nearly normal incidence (incidence angle 1º) with linearly polarized light in the scattering plane (shadowed plane in the Figure). The sample is rotated so that the narrow gap between the Py disk and the Au ring lies in the scattering plane (i.e., parallel to the electric field of the incident radiation). An out-of-plane magnetic field from an electromagnet saturates the magnetization of Py nanodisks. **b** Kerr rotation ($\theta_K$) and ellipticity ($\varepsilon_K$) and **c** MOA experimental spectra for NCRD and Py-DI structures. **d** Kerr rotation ($\theta_K$) and ellipticity ($\varepsilon_K$) and **e** MOA simulated spectra for NCRD and Py-DI structures. The dark orange arrows in panels **c** and **e** mark the amplification of the MOA for the NCRD with respect to Py-DI structures.

A clear difference can be observed both in shape and in intensity of the experimental MOA for the Py-DI and the NCRD structures (Fig. 3c). For the case of the Py-DI nanoantennas, the MOA spectrum displays the usual features of magnetoplasmonic nanostructures, with a maximum of the MOA (gray balls in Fig. 3c) in the spectral range were the dipolar LPR is observed, with the characteristic oscillating behavior of $\theta_K$ and $\varepsilon_K$ (solid dots in Fig. 3b). These lineshapes are well understood and known to arise from the interplay between the amplitude and phase of the LPR and MOLPR.[27] The fact that MOKE signals of the Py-DI sample can be clearly measured in the spectral range where the LPR is excited (500-900 nm) despite the fact that the nanoantennas cover only a minute fraction (≈1.2%) of the sample surface, is the result of the well-studied plasmon enhancement of the MOA. It is worth noting here that for NCRD structures only the Py disk is contributing to the MOKE signal, as any ***H***-dependent contribution from the Au ring portion of the nanocavity is not measurable for the weak ***H*** utilized in the experiment (indeed the Au-RI sample does not show any detectable MOKE signal). Therefore, the surface density of the magneto-optical active material is *exactly the same* in both the Py-DI and NCRD samples and the MOKE spectra can be compared side by side. The most striking result shown in Figs. 3b and 3c is the *additional* and approximately *one order of magnitude higher enhancement* of the MOA, $\theta_K$, and $\varepsilon_K$ on the NCRD nanocavities near 820 nm of wavelength (marked by a dark orange arrow in Fig. 3c) with respect to the Py-DI nanoantennas. The highly enhanced MOA at 820 nm can be appreciated by looking at the direct comparison of the measured $\theta_K$, and $\varepsilon_K$ signals at 820 nm as a function of the applied field ***H*** for the NCRD and Py-DI shown in Fig. S2. Experimental spectral lineshapes of the MOA, $\theta_K$, and $\varepsilon_K$ are excellently reproduced by simulations as it can be seen from comparison of Figs. 3b-c with Figs. 3d-e. Simulations in Fig. 3d predict an even larger enhancement (exceeding the 1-order of magnitude) of the MOA, $\theta_K$ and $\varepsilon_K$ for defect-free NCRD nanocavities. This remarkable additional amplification of the MOA is of outmost relevance in the view of applications in active nanophotonics and flat-optics.

In order to shed light on the underlying mechanism at work in the case of NCRD nanocavities at 820 nm wavelength, we first calculated the relative strength of the photoinduced electric dipole $p_O$ associated to the LPR and of the ***H***-activated electric dipole $p_{MO}$ associated to the MOLPR both in the Py-DI and in the Py nanodisk part of a NCRD nanocavity. The results of this comparative analysis are summarized in Fig. 4, which displays the computed 2D-maps of the surface charge ($\sigma$) generated by $p_O$ and $p_{MO}$ (see Methods and Fig. S3 for details). The most interesting result from the comparison of the maps in Fig. 4 is that, at 820 nm, the strength of both induced dipoles $p_O$ and $p_{MO}$, which is proportional to the maximum value of $|\sigma|$ ($|\sigma|_{max}$), in the Py disk inside the NCRD nanocavity is much larger (~1 order of magnitude at 820 nm) than that in the bare Py-DI nanoantenna at the same wavelength and even at full resonance at 550 nm.

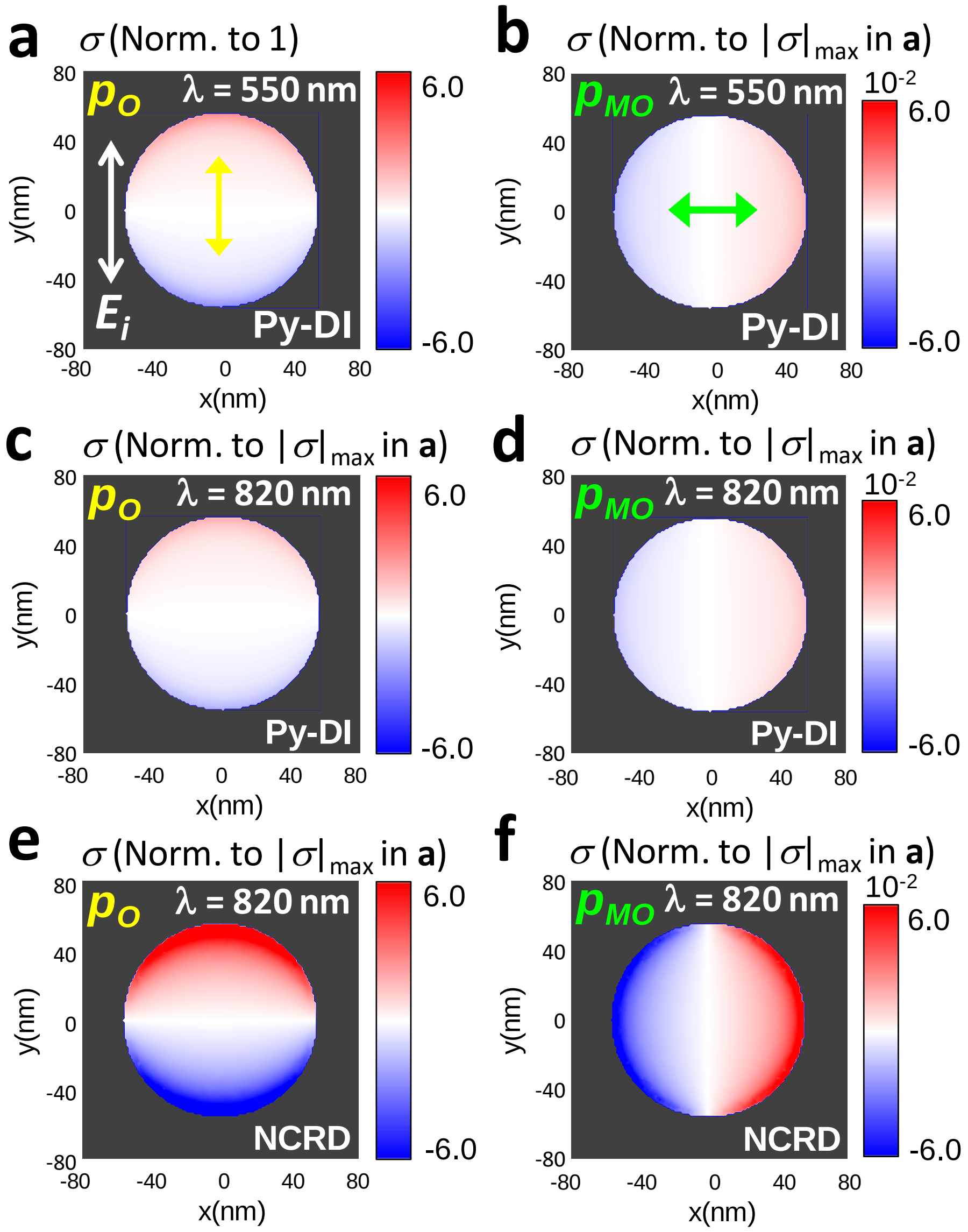

**Fig. 4 Relative strength of induced optical ($p_O$) and magneto-optical ($p_{MO}$) electric dipoles of the Py-disk in the NCRD structure and a standalone Py-disk.** 2D maps of the calculated surface charge distribution $\sigma$ at the interface between the structures and the substrate. The intensity of all maps is normalized to the maximum value of $|\sigma|$ ($|\sigma|_{max}$) produced by the optical dipole $p_O$ for the Py-DI at 550nm shown in panel **a.** The yellow arrow in **a** displays the direction of the optical dipole $p_O$. **b** displays the 2D map of normalized $\sigma$ produced by the magnetic field (***H***) induced electric dipole $p_{MO}$ (the green arrow shows its direction) at the same wavelength of 550 nm (details of the calculation given in Methods). **c** and **d** display the corresponding 2D maps of the normalized $\sigma$ for the Py-DI at the wavelength of 820 nm. **e** and **f** show the corresponding 2D maps of the normalized $\sigma$ for the NCRD at the wavelength of 820 nm (Fano resonance maximum). $E_i$ in panel **a** (equal to 1 V/m for all maps) shows the direction along which the incident light is polarized. The color scale range of the 2D maps in each column is kept fixed to better highlight the relative size of the induced electric dipoles. The range in each column is chosen as the ratio between $|\sigma|_{max}$ in the NCRD at 820 nm and in the Py-DI at 550 nm corresponding to $p_O$ (first column) and $p_{MO}$ (second column). The σ maps at 820 nm for the Py nanodisk inside the NCRD showed the appearance of a weak quadrupolar mode ($\boldsymbol{S_4}$) superimposed to the intense and dominating dipolar modes $\boldsymbol{p_O}$ and $\boldsymbol{p_{MO}}$ (caused by hybridization of $\boldsymbol{p_O}$ with the mode $\boldsymbol{S_6}$ of the nanoring; see Fig. S3 and its caption, and Supplementary Video_V1). Since such weak quadrupolar mode is non-radiative, and thus not relevant for the discussion, its contribution was removed in Fig. 4 to ease the relative comparison of induced dipoles strength (see discussion in the caption of Fig. S3).

This remarkable enhancement occurs despite the fact that the Py disk is not driven at full resonance (the strength of $p_O$ and $p_{MO}$ is reduced by a 20% at 820 nm as we evaluated by comparing Figs. 4a with 4c or equivalently 4b with 4d). The explanation of this so prominent enhancement comes from the detailed analysis of the plasmonic coupling-induced electrodynamics in the Py disk inside the NCRD nanocavity. This is performed by simulated comparative monitoring of the time evolution of the charge density induced by the electric field of the incident light in the NCRD cavity and standalone Au-RI. At each time, the surface charge map of the Au-RI (multiplied by a factor, see Fig. 2c) is subtracted from that of the of the NCRD nanocavity. This allows the identification of the modes, and their symmetry, that hybridize in the NCDR nanocavity. The full analysis is shown in Fig. S4 and the key results are summarized in Fig. 5a.

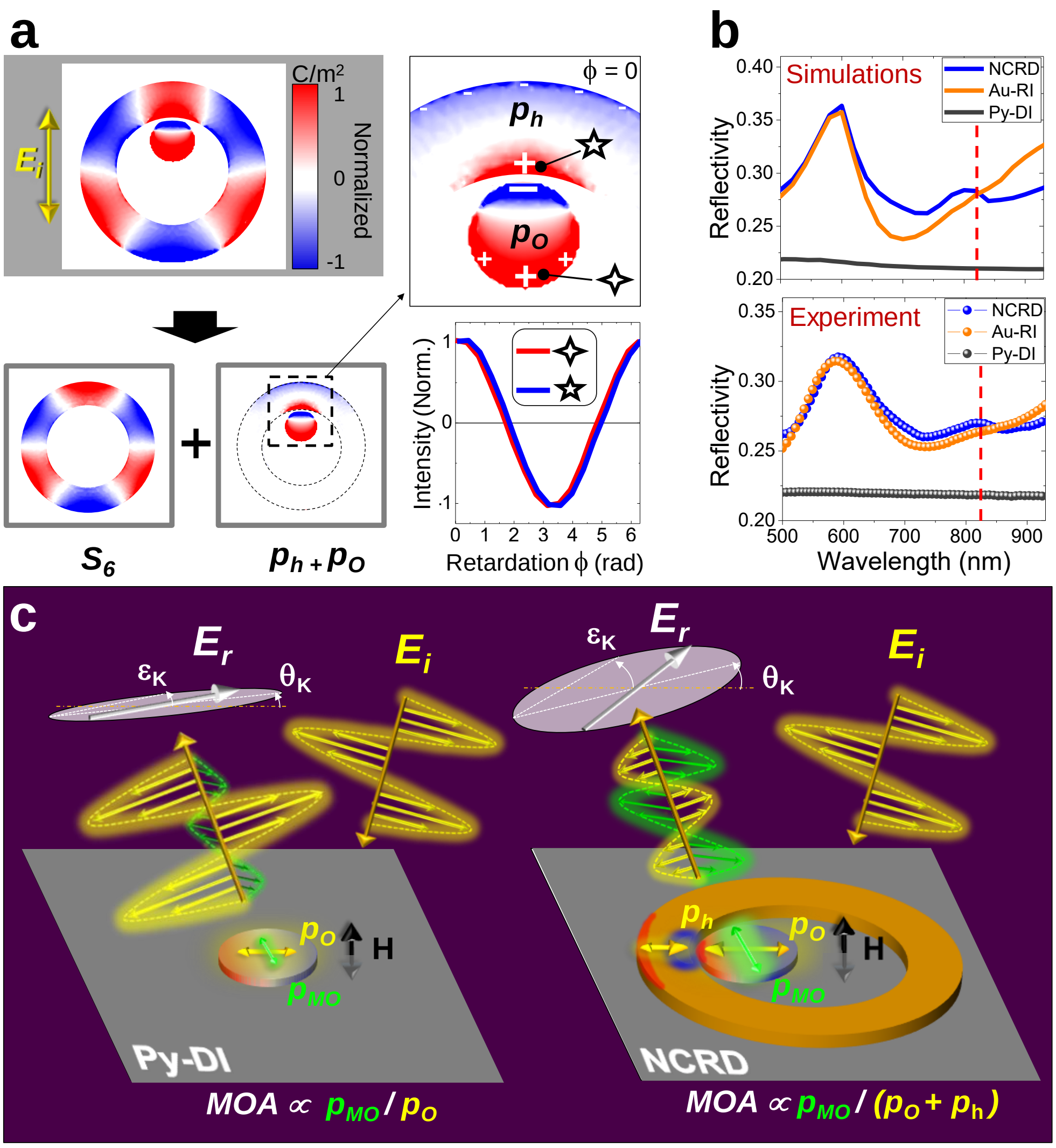

**Fig. 5 Illustration of the resonant mechanism leading to the large magneto-optical enhancement in the NCRD nanocavity at the Fano resonance maximum**. **a** Separation of the hybrid mode into its dipolar ($\boldsymbol{p_O}$ and $\boldsymbol{p_h}$) and multipolar ($\boldsymbol{S_6}$) components (these are the surface charge maps corresponding to phase $\phi_0$ in the first column of Fig. S3). The panel shows also the time evolution (retardation $\phi$ - $\phi_0$) of the surface charge in the points marked by the 4- and 5-pointed stars. **b** Calculated and measured optical reflectivity spectra for the NCRD, Au-RI and Py-DI structures. Red dashed lines mark the wavelength of 820 nm showing that the reflectivity of the NCRD is only marginally higher than that of the Au-RI. Reflectivity of both NCRD and Au-RI are approximately 25% higher than that of Py-DI. **c** A sketch of the electrodynamics of the Py disk generating an electric dipole ($\boldsymbol{p_O}$) triggered by the electric field $\boldsymbol{E_i}$ of an incident linearly polarized electromagnetic radiation and a magneto-optically activated electric dipole ($\boldsymbol{p_{MO}}$) by a magnetic field $\boldsymbol{H}$. $\boldsymbol{p_O}$ and $\boldsymbol{p_{MO}}$ of the Py nanoantenna inside the NRCD nanocavity are enhanced (by a factor of ~5)

with respect to a bare Py disk (Py-DI) by hybridization with the mode $\boldsymbol{S_6}$ of the Au-RI that results in the introduction of a dipolar mode, $\boldsymbol{p_h}$, in the $\boldsymbol{S_6}$ mode. This is qualitatively depicted by the relative size of electric dipoles $\boldsymbol{p_O}$ and $\boldsymbol{p_{MO}}$ in the Py-DI, and in the NCRD. In the NCRD nanocavity, hybridization generates a bonding dipolar mode $\boldsymbol{p_O} + \boldsymbol{p_h}$. In the Py-DI system both $\boldsymbol{p_O}$ and $\boldsymbol{p_{MO}}$ are generated by *radiant* (bright) LPR modes and the resulting $\boldsymbol{H}$-induced polarization change in the reflected radiation, $\boldsymbol{E_r}$, is determined by their ratio (MOA $\propto |\boldsymbol{p_{MO}}|/|\boldsymbol{p_O}|$, $\theta_K \propto \mathrm{Re}[\boldsymbol{p_{MO}}/\boldsymbol{p_O}]$, and $\varepsilon_K \propto \mathrm{Im}[\boldsymbol{p_{MO}}/\boldsymbol{p_O}]$). The large enhancement of the $\boldsymbol{H}$-induced polarization change in the NCDR system is a consequence of the *low-radiant* character of the bonding dipolar mode ($\boldsymbol{p_O} + \boldsymbol{p_h}$), whilst $\boldsymbol{p_{MO}}$ maintains its enhancement and *radiant* character.

Fig. 5a clearly shows that the electrodynamics of the NCRD at 820 nm (see also Supplementary Video_V1) is determined by the Fano resonance resulting from hybridization between the bright dipolar mode $\boldsymbol{p_O}$ of the Py nanodisk with the $\boldsymbol{S_6}$ dark mode of the Au ring forming the nanocavity $\boldsymbol{p_O} + \boldsymbol{S_6}$ mode (referred to as multipolar Fano or octupolar Fano resonance in the literature[34]). The strong and localized near-field coupling between the two parent modes introduces an additional dipolar mode, $\boldsymbol{p_h}$, into the ring nanocavity (marked on Fig. 5a). Therefore, overall at the Fano resonance the electrodynamics can be decomposed in a coupled mode $\boldsymbol{p_c} = \boldsymbol{p_O} + \boldsymbol{p_h}$ and the dark, i.e. non-radiating, $\boldsymbol{S_6}$ mode. The analysis of the relative phase of the electric dipoles $p_h$ and $p_O$ associated to the coupled mode $\boldsymbol{p_c}$ reveals that, at the wavelength of 820 nm, the two dipoles oscillate in phase. Thereby, at this wavelength the Fano resonance displays its maximum due to constructive interference. In this situation, the coupled mode $\boldsymbol{p_c}$ has a bonding character, which results in the marked enhancement of the local dipole $p_O$ of the Py disk. This enhancement is transferred to the $\boldsymbol{H}$-activated dipolar mode $\boldsymbol{p_{MO}}$ via the MO coupling in Py (compare Figs. 4b and 4f). Noteworthy, simulations show also that the mode $\boldsymbol{p_{MO}}$ does not show a significant direct hybridization with any mode of the ring nanoresonator.

Although intriguing, *per se* the simultaneous enhancement of the electric dipoles $p_O$ and $p_{MO}$ does not explain the surprisingly large enhancement of the MOA produced by the NCRD nanocavity at 820 nm. Indeed, if both dipoles were bright, i.e., radiant, the MOA should be quite similar for the three cases analyzed in Fig. 4. The answer to our puzzle is again contained in Fig. 5a: the uneven charge distribution in the bonding-coupled $\boldsymbol{p_c}$ mode (dominating tri-polar character, see Fig. 5a) makes this mode, and thus the hybrid multipolar Fano resonance mode $\boldsymbol{p_O + S_6}$, *low-radiant*. The *low-radiant* character of the Fano resonance mode at 820 nm is confirmed by Fig. 5b where we plot the calculated and measured reflectivity spectra $E_r/E_i$, with $E_i$ and $E_r$ being the incident and reflected electric far fields. Fig.5b clearly shows that the excitation of the multipolar Fano resonance mode $\boldsymbol{p_O + S_6}$ enhances the reflectivity of the NCRD only marginally in respect to the Au-RI. Figure 5b shows also that the reflectivity of the NCRD at 820 nm is only 25% larger than that of the Py-DI (see also Fig. S5a that features the ratio between the reflectivity of the NCRD and Py-DI). This indicates that the dip at around 820 nm of wavelength in calculated and measured transmittance spectra (Figs. 2a and 2b) is predominantly due to radiation absorption, as illustrated by the comparison between the absorption cross section of the NCRD and Au-RI in Fig. S5b. The physical mechanism behind the large MOA amplification achieved in the NCRD, much larger than that of the parent magnetoplasmonic Py-DI structure, is now clear and is summarized in pictorial form in the cartoons shown in Fig. 5c. The cartoons depict schematically how, at 820 nm of wavelength, the hybridization between $\boldsymbol{p_O}$ and $\boldsymbol{S_6}$ in the NCRD nanocavity leads to the excitation of the *low-radiant* bonding mode $\boldsymbol{p_c}$ (the $\boldsymbol{S_6}$ mode of the Fano resonance is not shown in Figure for clarity) while the $\boldsymbol{H}$-activated dipolar mode $\boldsymbol{p_{MO}}$, which is only weakly interacting with the Au ring, inherits the large enhancement from $\boldsymbol{p_O}$ while preserving its *radiant* character. As a consequence, the unique electrodynamics produced by the hybridization in the nanocavity occurring at around 820 nm generates a strongly enhanced $\boldsymbol{H}$-activated bright electric dipole $p_{MO}$, *without paying the price* of a parallel increase of the re-emitted radiation with primary polarization, thanks to the *low-radiant* character of the driving hybrid multipolar Fano mode $\boldsymbol{p_O + S_6}$. The result is a large MOA amplification in the NCRD nanocavity, ~1-order of magnitude larger than that of the parent magnetoplasmonic Py-DI structure (Figs. 3c and 3e). Further corroborating the uniqueness of the electrodynamics occurring in the NCRD nanocavity at 820 nm (see Supplementary Video_V1), we can note that even stronger hybridization of the LPR of the Py disk with the intense antibonding mode of the Au ring (so-called superradiant mode[34]) in the spectral range of 500-600 nm (Fig. 2c) does not produce any enhancement of the MOA of the NCRD with respect to the Py-DI. Remarkably, it is rather the opposite: the large enhancement of the reemitted radiation by this superradiant mode in the Au ring results in a suppression of the MO response at 600 nm as it appears from the spectra in Figs. 3c and 3e. Finally, and as a further proof, we report in Fig. S6 the optical (panel b) and MO (panel c) responses of a concentric ring/disk (CRD) nanocavity with sizes of the constituent parts nominally identical to those of the NCRD (see SEM in panel a). For such CRD nanocavity the MO response at 820 nm should be the same as that of the Py-DI structures (see surface charge maps in panel d of Fig. S6) given the geometrical symmetry of the structure. The experimental MOA of the CRD sample is indeed only weakly modified with respect to the Py-DI, because of a non-perfect concentricity of the nanocavity.

## Conclusions

We have demonstrated that high-order multi-polar dark plasmon resonances in magnetoplasmonic nanocavities can be utilized to achieve unprecedented enhancement of the magneto-activated optical response, beyond the present limitations of magnetoplasmonic nanoantennas, enabling a far more efficient active control of the light polarization under weak magnetic fields. The superior behavior of geometrical symmetry broken magnetoplasmonic nanocavities, as compared to corresponding nanoantennas, is explained by the generation of largely enhanced magnetic-field-induced *radiant* dipole in the magnetoplasmonic nanoantenna driven by a hybrid *low-radiant* multipolar Fano resonance mode. Therefore, in this novel design, a large enhancement of the magneto-optical response, i.e., the magneto-activated electrical dipole inducing the light polarization modification, is achieved without a significant increase of the pure optical response thanks to the *low-radiant* character of the hybrid mode. As a result, in the NCRD magnetoplasmonic nanocavity the MOA is additionally amplified by ~1-order of magnitude with respect to the parent Py-DI structure. The novel concept unveiled here opens a fresh path towards applications of magnetoplasmonics to a variety of fields ranging from flat and active nanophotonics to sensing. Therefore, this exploratory work should catalyze future research. Tuning of dark and bright plasmon modes can be achieved by varying the design and the materials to boost both plasmonics (e.g. using silver instead of gold) and intrinsic MO activity (e.g., employing multilayers Au/Co), as well as tuning the relative spectral position and sharpness of the dark and dipolar modes, and thus of the Fano resonance line shape and intensity. Finally, this mechanism might have a huge impact on forthcoming photonic nanotechnologies based on plasmon-mediated local enhanced manipulation of electronic spin-currents opening excellent perspectives in disclosing novel opto-electronic phenomena.

## Acknowledgements

This work was supported by the Spanish Ministry of Economy, Industry and Competitiveness under the Maria de Maeztu Units of Excellence Programme - MDM-2016-0618 and from the European Union under the Project H2020 FETOPEN-2016-2017 "FEMTOTERABYTE" (Project n. 737093).

**Author Contributions**

PV devised the concept; ALO performed nanofabrication, optical, and magneto-optical experimental measurements; AC performed EELS measurements; MZH, NM and MP performed numerical simulations and MG analytical calculations. PV and ALO wrote the manuscript and all authors contributed to the discussions.

**Competing interests**

The authors declare no competing interests.

**Additional information**

Supplementary information is available for this paper at…

Correspondence and requests for materials should be addressed to ALO and PV

**Data availability statement**

The authors declare that all data supporting the findings of this study are available within the paper and its supplementary information files.

## Methods

**Fabrication**. The samples were fabricated by electron-beam (e-beam) lithography and lift-off procedure. Initially, a 5 nm thick Ti layer was e-beam evaporated (evaporation rate: 0.4 Å/s) onto a cleaned 10 mm x 10 mm Pyrex substrate, as an adhesion layer and as a metallic coating for avoiding charging effects during e-beam lithography. The NCRD structures have been grown by a sequential two-step procedure. Firstly, the Py disk was prepared by spin-coating a positive resist (ZEP520A-7) onto the substrate at 4000 rpm for 60 s. The resist was exposed by a 20 KV electron beam inside a RAITH eLine system. Exposition time was adjusted according to the dot size. After developing the exposed resist (with ZED-N50 developer), a 40 nm thick Py layer was thermally evaporated (evaporation rate: 0.8 Å/s). Finally, the lift-off was carried out by dipping the samples in the proper solvent (ZDMAC). Eventually, the process has been repeated for fabricating the Au ring around the disks.

**Structural, optic, electron and atomic force microscopy and magneto-optic characterization.** Scanning electron microscopy images have been recorded in an eSEM-FEI QuantaTM 250 instrument operating at an accelerating voltage of 10 kV. The transmittance spectra were taken in the wavelength range 400-1600 nm using the Pyrex glass as the background signal. EELS and scanning transmission electron microscopy (STEM) images have been acquired using a TitanG2 60-300 (FEI, Netherlands) operating at 80 kV in monochromatic mode (energy resolution ~80meV as FWHM of ZLP). Atomic force microscopy images were acquired in air under ambient conditions using a Nano Observer system (CSI Company, France). Measurements were made in tapping mode using a NCHV-A (Bruker) tip with spring constant k = 40 N/m and a resonance frequency of 320 kHz. Magneto-optical Kerr effect measurements were conducted with a Kerr spectrometer. A schematic of the setup and geometry utilized in the experiment are shown in Fig. S1. The setup consisted of a broadband supercontinuum laser (SuperK Extreme EXR-15 from NKT Photonics), polarizing and focusing optics, a photoelastic modulator (Hinds Instruments II/FS42A), and a Si-photodetector (Thorlabs PDA 36A-EC). The wavelength of the laser was tuned between 500 nm and 1200 nm. We used linear polarized light at normal incidence (polar Kerr configuration). During measurements, a ±700 mT magnetic field from an electromagnet switched the magnetization of the Py nanodisks between the two perpendicular directions (field for magnetic saturation is ±350 mT, see Fig. S2). Two lock-in amplifiers were used to filter the signal at the modulation frequency (42 kHz) and at twice the modulation frequency in order to retrieve Kerr ellipticity ($\varepsilon_K$) and rotation ($\theta_K$) angles simultaneously.[47] The limit of detection of our MOKE setup, i.e. the noise level, is of 2 μradians (standard deviation), namely ~2 orders of magnitude smaller than the smallest signal measured in the here reported experiment. Therefore, error bars are smaller than the size of the symbols utilized in the plots (Figs. 3b and 3c).

**Simulations**. **Electromagnetic simulations:** 3D electrodynamical calculations of the optical transmittance and the surface charge density maps were performed adopting the Finite Element Method implemented in the commercial COMSOL Multiphysics software[48] using the RF module in the frequency domain. The experimental structures (Py-DI, Au-RI and NCRD) were modeled as arrays using standard ports formulation and periodic boundary conditions. Hence, the physical domains resulted to be placed in regular square array arrangements with a pitch of 800 nm along both the in-plane axes and influenced by linearly polarized light at normal incidence. We used air, $n_{air}$=1.0, for the incoming light environment, a substrate with a refractive index n=1.5 (mimicking Pyrex) and Au dielectric optical functions from Johnson and Christy.[49] For the magneto-plasmonic structure (Py-DI) we consider a non-diagonal dielectric tensor medium in which the non-zero off-diagonal elements depend on the applied magnetic field, the orientation of the geometry and the polarization of the incoming light. For our case, i.e. light reflected through the sample with an applied magnetic field perpendicular to the surface of the sample (polar Kerr configuration), the dielectric tensor for the Py-DI in terms of the diagonal $\varepsilon_d(\omega)$ and off-diagonal $\varepsilon_{od}(\omega)$ terms adopts the form

$$\varepsilon(\omega) = \begin{pmatrix} \varepsilon_d(\omega) & \mp\varepsilon_{od}(\omega) & 0 \\ \pm\varepsilon_{od}(\omega) & \varepsilon_d(\omega) & 0 \\ 0 & 0 & \varepsilon_d(\omega) \end{pmatrix},$$

where $\varepsilon_d(\omega)$ and $\varepsilon_{od}(\omega)$ for Ni are taken from Ref.[50]. We used dielectric properties of Ni as representative of Py since the two materials have almost identical optical and magneto-optical properties and the dielectric tensor constants of the former are available over a larger spectral range. The inversion of the sign in the off-diagonal elements mimics the effect of the reversal of the static magnetic field from + *H* to -*H*. All domains were meshed by using tetrahedral elements where the maximum mesh element size was kept below λ/10, where λ is the wavelength of the incident light. For the elements corresponding to both the ring and the disk domains, the size was ten times finer than the biggest element size (verified to properly resolve the considered structures).

The 2D-maps of the surface charge density (Fig. 2c, Fig. 5a, Fig. S3, Fig. S4, and Fig. S6d) used for the relative comparison of the dipoles strength were obtained by plotting computed σ = **P·n** in a plane parallel to the interface between the structures and the substrate (**P** is the electric dipole per unit surface, i.e. the electric polarization, and **n** is the normal to the surface). To extract the 2D-maps of σ produced by the MOLPR in Fig. 4 and of |σ| in Fig. S4d, we subtracted the COMSOL calculated **P·n** at +***H*** and -***H***. In this respect, we mention here that the σ maps at 820 nm for the Py nanodisk for the NCRD showed the appearance of a weak quadrupolar mode ($\boldsymbol{S_4}$) superimposed to the intense and dominating dipolar modes $\boldsymbol{p_O}$ and $\boldsymbol{p_{MO}}$ (caused by hybridization with the mode $\boldsymbol{S_6}$ of the nanoring). Since such weak quadrupolar mode is non-radiative, and thus not relevant for the discussion, its contribution was removed to ease the relative comparison of induced dipoles strength.

**Normal modes simulations:** For the calculation of the Au-RI plasmonic normal modes we adapted the geometry of our system to the efficient finite-element solver (QNMEig) developed in COMSOL Multiphysics by Yan *et. al*,[46] which uses a Quasi-normal mode (QNM)-expansion formalism to compute the resonance modes of an absorptive and dispersive plasmonic nanoresonator solving a standard linear eigenvalue problem derived from Maxwell's equations. A key quantity retrieved by the formalism is the electric polarization **P**, from which the surface charge density can be calculated.

## Supplementary Information

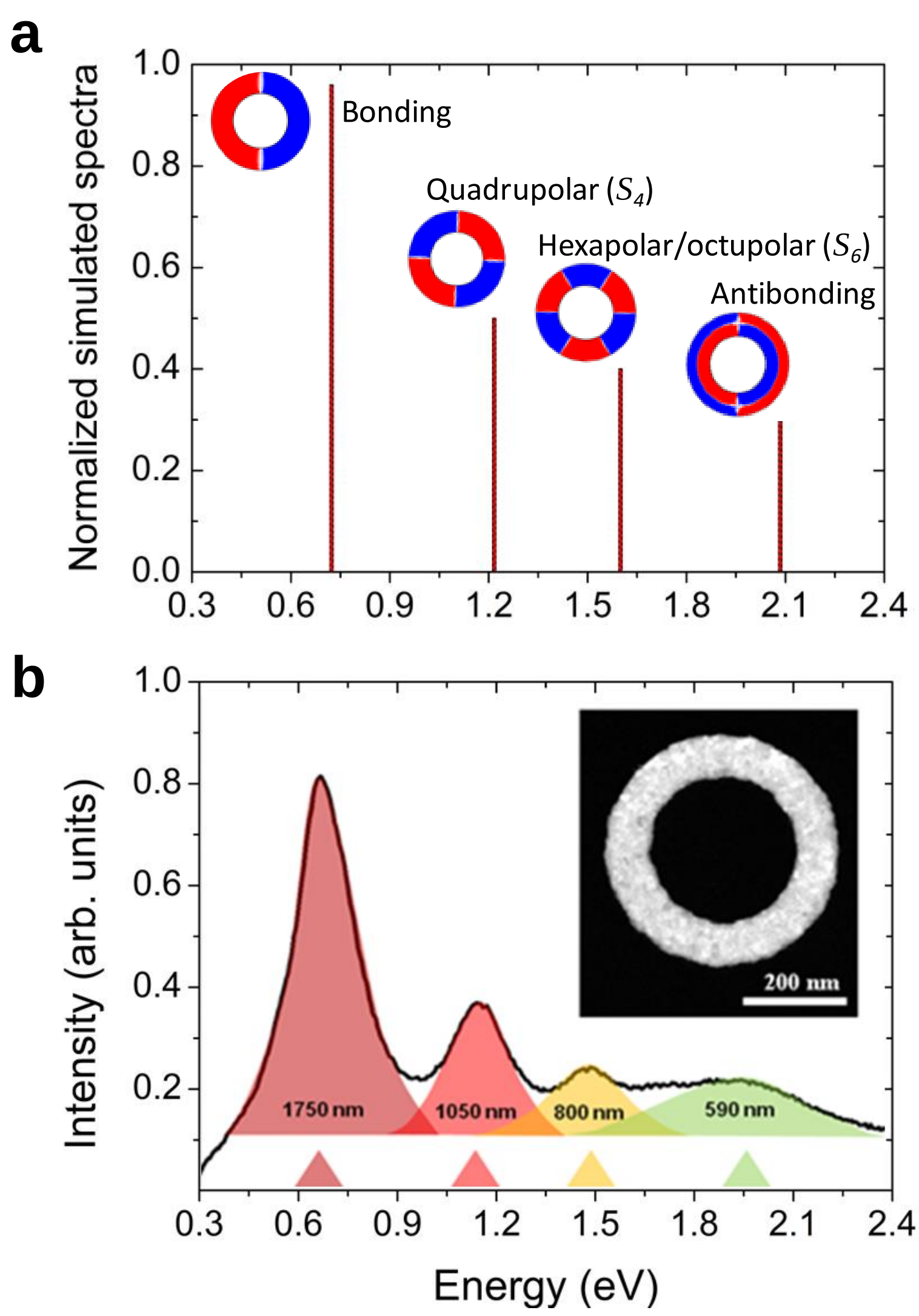

**Fig. S1 Plasmonic modes of Au nanorings in the Vis-IR spectral region. a** Theoretical reconstruction of the symmetry of the surface charge density maps and simulated spectrum for an Au ring showing the peaks as returned by the (QNM)-expansion formalism for defect-free structures.[1] **b** EELS spectrum obtained for an Au ring grown on a $SiN_2$ membrane and the respective scanning STEM image.

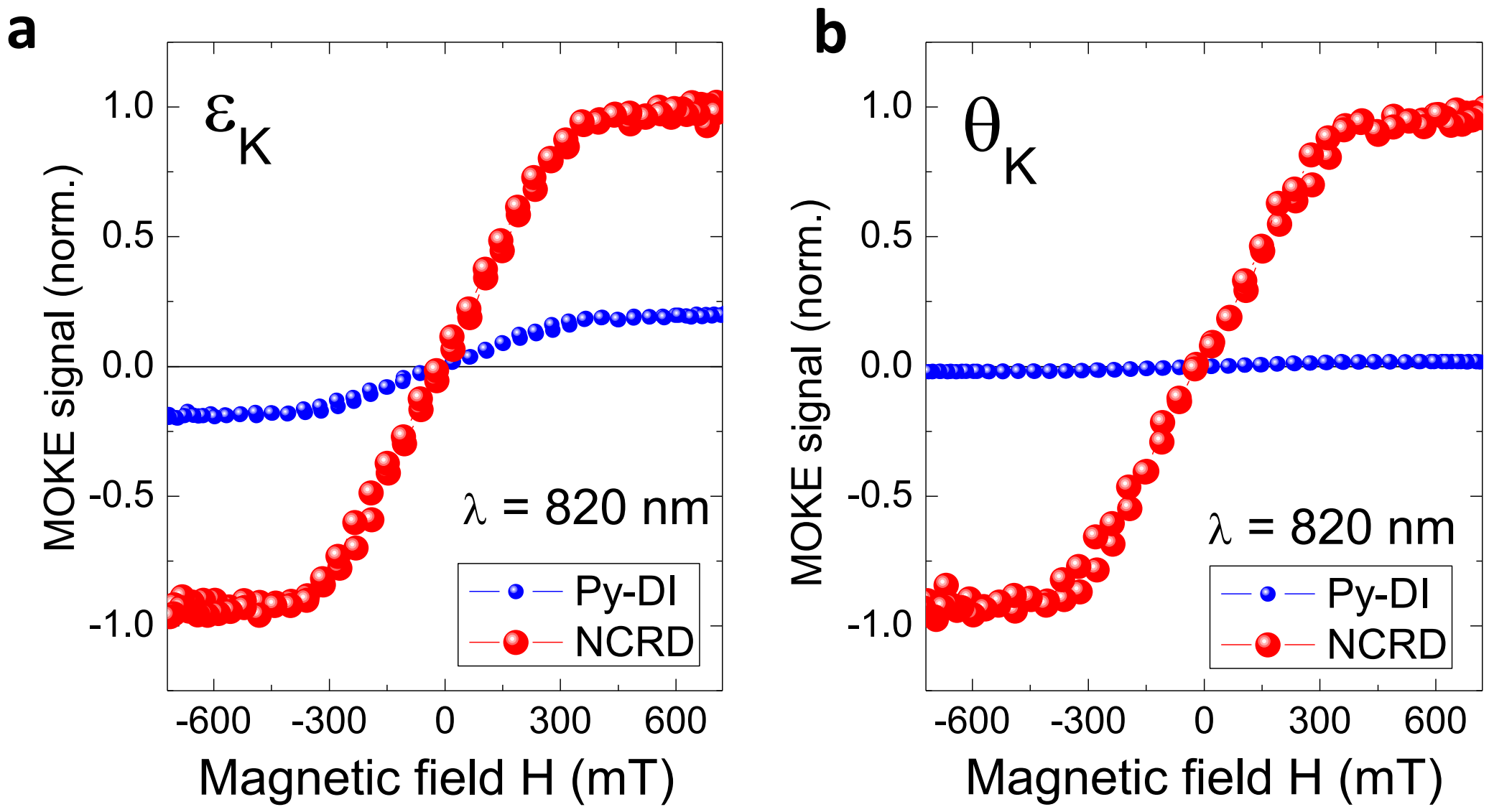

**Fig. S2 measured magnetic field dependent Kerr rotation and ellipticity signals**. **a** $\varepsilon_K$ and **b** $\theta_K$ signals measured at a wavelength of 820 nm as a function of the applied field H for the NCRD and Py-DI. The signals have been normalized to the saturation value (high |H|) for the NCRD.

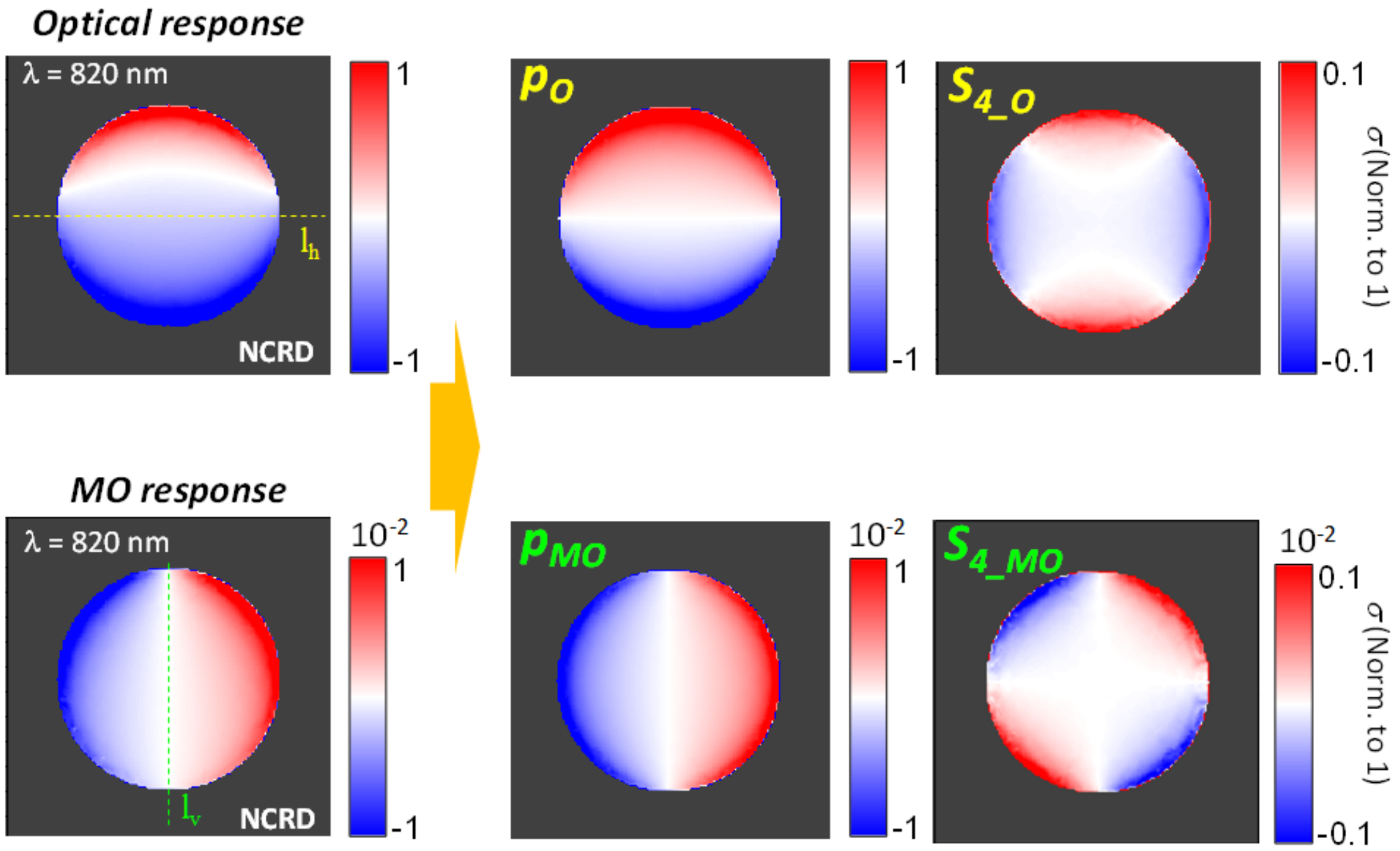

**Fig. S3 Multipolar analysis of the optical and magneto-optical response of the Py disk in the NCRD at 820 nm for the NCRD.** 2D maps of COMSOL calculated surface charge density $\sigma$ produced by the optical and magnetic-optical activity responses of the Py disk inside the NCRD at the interface between structure and substrate. Intensity is normalized to the maximum value of $|\sigma|$ produced by the optical response (topmost left panel). Imposing a top-bottom and left-right anti-symmetry with respect to the dashed lines $l_h$ and $l_v$ to the spatial distribution of σ resulting from the optical and magneto-optical responses, respectively, one singles out the optical ($p_O$) and magnetic-activated ($p_{MO}$) electric dipoles. The 10-times-smaller residuals of these symmetry operations display a $\boldsymbol{S_4}$ (quadrupolar) symmetry. Interesting, optical and magneto-optical $\boldsymbol{S_4}$ modes ($\boldsymbol{S_{4_O}}$ and $\boldsymbol{S_{4_MO}}$) are rotated by 45 degrees one respect to the other. The $\boldsymbol{S_4}$ modes are π/4 radians out of phase with respect to their respective dipolar mode.

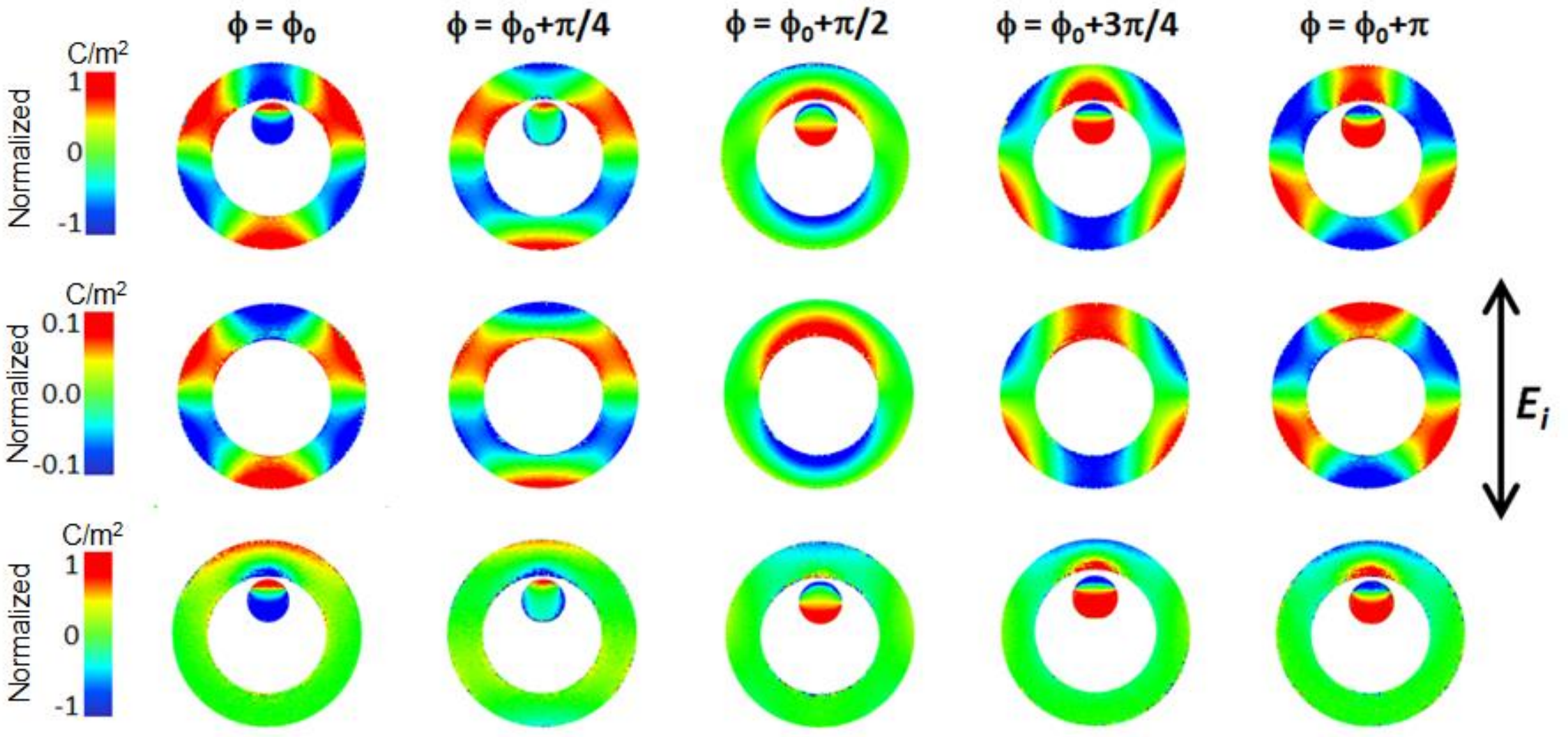

**Fig. S4 Surface charge distribution maps at 820 nm for the NCRD and Au-RI in function of time.** The topmost sequence shows the NCRD surface charge density maps, normalized to 1, at 820nm and at different values of the phase delay $\phi$ (i.e., as a function of time) with respect to an arbitrarily chosen initial phase $\phi_0$ (i.e., time $t_0$) over half-period ($\phi-\phi_0 \in [0\text{-}\pi]$). The sequence in the middle displays the surface charge density (normalized to the values of the NCRD) map evolution at 820nm for the bare Au-RI. The bottom-most sequence shows the evolution of the difference between the topmost and middle (multiplied by a factor 10) sequences. Simulations were carried out using linearly polarized electromagnetic radiation as indicated by the black arrow ($E_i$ = 1V/m).

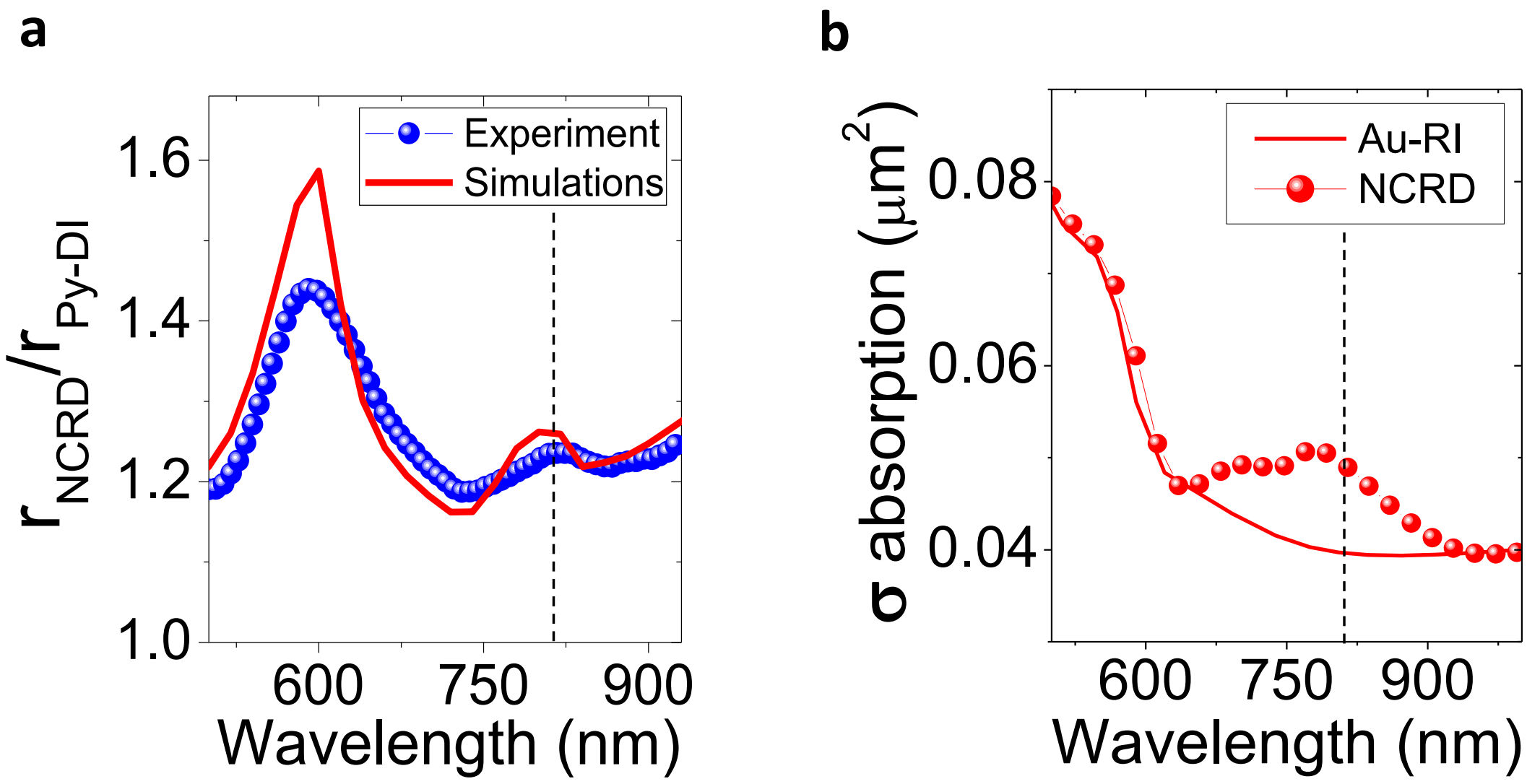

**Fig. S5 Relative reflectivity of NCRD with respect to Py-DI and absorption cross section with respect to Au-RI. a** Experimental and simulated spectral dependence of the ratio between the reflectivity of the NCRD ($r_{NCRD}$) and that of the Py-DI ($r_{Py\text{-}DI}$). In our experiment, the reflectance signal ($R = r^2$) is recorded simultaneously to the MOKE one and utilized to generate the Kerr rotation ($\theta_K$) and ellipticity ($\varepsilon_K$) spectra of Fig. 3. **b** Calculated optical absorption cross section spectra for the NCRD and Au_RI (red symbols and line, respectively). The dashed gray lines mark the 820 nm wavelength.

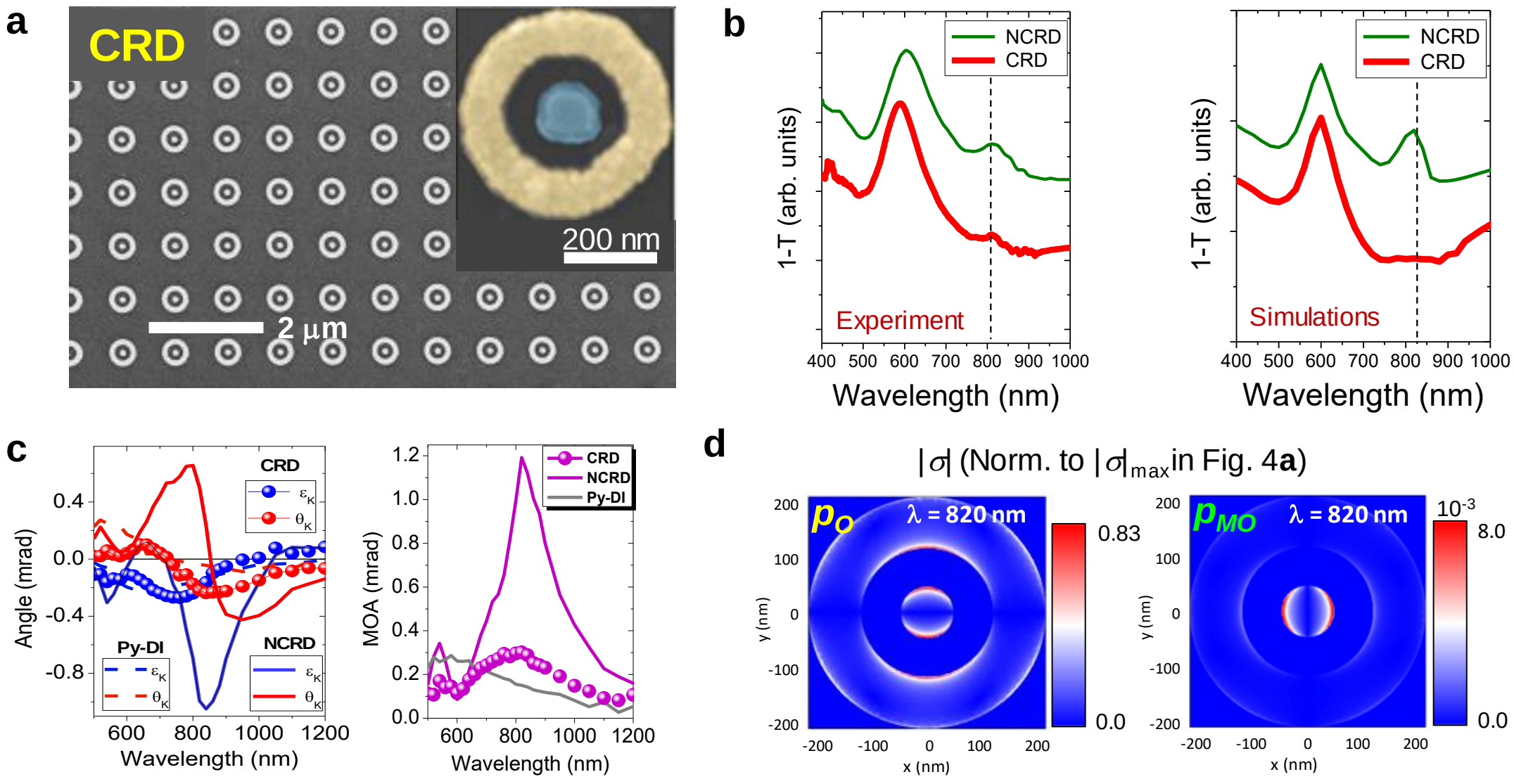

**Fig. S6 SEM and MOKE of CDR cavities. a** Scanning electron microcopy images of the array of CRD nanocavities. **b** Experimental (left) and simulated (right) transmittance spectra for the CRD and NCRD nanocavities. **c** Kerr rotation ($\theta_K$) and ellipticity ($\varepsilon_K$) and MOA experimental spectra for CRD (solid symbols) together with those for NCRD and Py-DI structures (lines). **d** 2D maps of COMSOL calculated $|\sigma|$ produced by the optical ($p_O$) and magnetic-activated ($p_{MO}$) electric dipoles at the interface between structures and substrate. Intensity is normalized to the maximum value of $|\sigma|$ produced by the optical dipole $p_O$ for the Py-DI at 550nm shown in Fig. 4a.